\documentclass[%
reprint,
amsmath,amssymb,
aps,
]{revtex4-1}

\usepackage{graphicx}% Include figure files
\usepackage{dcolumn}% Align table columns on decimal point
\usepackage{bm}% bold math
\begin{document}
	%\preprint{APS/123-QED}
	
	\title{Multiple Exclusion Statistics}% Force line breaks with \\
	
	\author{Julian J. Riccardo}
	\email{Corresponding author. jjriccardo@unsl.edu.ar}
	% \altaffiliation[Also at ]{.}%Lines break automatically or can be forced with \\
	\author{Jose L. Riccardo, Antonio J. Ramirez-Pastor, Marcelo P. Pasinetti}%
	\affiliation{%
		Departamento de F\'{\i}sica, Instituto de F\'{\i}sica
Aplicada, Universidad Nacional de San Luis-CONICET, Ej\'ercito de
los Andes 950, D5700BWS, San  Luis, Argentina.
	}%
	
	
	%\affiliation{
	%\affiliation{
	%\affiliation{%
	
	
	\date{\today}% It is always \today, today,
	
	\begin{abstract}
		A new distribution for systems of particles in equilibrium obeying exclusion of correlated states is presented following the Haldane's state counting. It relies upon an ansatz to deal with the multiple exclusion that takes place when the states accessible to single particles are spatially correlated and it can be simultaneously excluded by more than one particle. The Haldane's statistics and Wu's distribution are recovered in the limit of non-correlated states of the multiple exclusion statistics.  In addition, an exclusion spectrum function $\mathcal{G}(n)$  is introduced to account for the dependence of the state exclusion on the occupation-number $n$. Results of thermodynamics and state occupation are shown for ideal lattice gases of linear particles of size $k$ ($k$-mers) where multiple exclusion occurs. Remarkable agreement is found with Grand-Canonical Monte Carlo simulations from $k$=2 to 10 where multiple exclusion dominates as $k$ increases. %The state-counting is also formally valid for fractional-exclusion particles.

		%\begin{description}
		%\end{description}
	\end{abstract}
	
	\maketitle
	
	%\tableofcontents
	
%\section{\label{sec:level1}Introduction}
	Quantum fractional statistics has drawn considerable interest in condensed matter physics since the early theoretical contributions \cite{leinaas1977theory,wilczek1982quantum,wilczek1982magnetic,halperin1984statistics,haldane1991fractional,wu1984general,wu1984multiparticle} and because of its ability to describe physical phenomena such as  fractional quantum Hall effect \cite{laughlin1983anomalous,laughlin1983quantized,halperin1984statistics}, spinor excitations in quantum antiferromagnets \cite{anderson1987resonating, haldane1991spinon}, high-temperature superconductivity \cite{laughlin1988rb}, quantum systems in low dimensions \cite{batchelor2006one,paredes2004tonks,kinoshita2004observation,jacqmin2011subpoissonian} and, more recently, its implications in the field of cosmology and dark matter.
	
	Concerning the quantum physics of strongly interacting many-particle systems, in a seminal work, Haldane \cite{haldane1991fractional} introduced the Quantum Fractional Statistics (FE) and the definition of the statistical exclusion parameter $g$, $0\leq g\leq 1$, being the Bose-Einstein (BE) and Fermi-Dirac (FD) the boundary statistics for $g=0$ and $g=1$, respectively. Later  Wu \cite{wu1994statistical} derived the statistical distribution for an ideal gas of fractional-statistic particles. These papers were a major contribution to describe quantum systems in one and two dimensions like anyons in a strong magnetic field in the lowest Landau level \cite {wilczek1990fractional} and excitations in pure Laughlin liquids \cite{laughlin1983anomalous, arovas1984quantumhalleffect, camino2005realization}.
	
	On the other hand, classical statistical mechanics of interacting large particles of arbitrary size and shape is a relevant problem since it is a major challenge to properly account for the generally complex entropic contribution to the free energy. Many physical systems, ranging from small polyatomics, alkanes, to protein adlayers, resemble these characteristics. The multisite occupancy problem has been addressed since long ago by the approximations of Flory-Huggins \cite{flory1942thermodynamics,huggins1942some,huggins1942thermodynamic,huggins1942viscosity} for binary solutions, lattice gases of particles of arbitrary size and shape made of a number $k$ of linked units ($k$-mers) \cite{dimarzio1961statistics} and it has been referred as the prototype of the lattice problem \cite{lieb1974exactly}. Among the motivations we can also mention Cooper and vortex pairs modelling \cite{cooper1956bound, kosterlitz1972long}, clusters diffusion on regular surfaces \cite{tsong1980migration,lin1990diffusion} and  thermodynamics of polyatomic adlayers \cite{paserba2001,strange2016,lopatina2018}, which represents a current open problem in statistical physics of gas-solid interfaces.
	The FE and Wu's distribution were already reinterpreted in the domain $g > 1$ to model the thermodynamics of linear $k$-mers ideal lattice gases behaving statistically like "superfermions" \cite{riccardo2004fractional} and resulting in the exact one-dimensional (1D) solution for $g=k$ \cite{ramirez1999statistical}. As shown later, in 1D it does not arise effective correlations between states, however it does in two or higher dimensions as considered here.
	
	This work addresses the statistical mechanics of identical particles in equilibrium occupying a set of spatially correlated states and obeying statistical exclusion in a confined region of the space. We refer as multiple exclusion the fact that, because of spatial correlations, the states accessible to single-particles can be simultaneously excluded by more than one particle in the system and it is not related to mutual exclusion as clearly defined by Haldane and Wu \cite{haldane1991fractional,wu1994statistical} to refer to exclusion statistics between different species within a space region.
	
	A classical realization of multiple exclusion phenomena are the physical models of lattice gases of $k$-mers.
	
In what follows, we develop a statistics for systems of many particles with state exclusion between spatially correlated states, which reduces to Haldane-Wu's FE for statistically independent states (constant exclusion $g$) and, correspondingly, to the FD and BE ones.
	Let us consider a system of volume $V$ containing $N$ identical particles having $G$ states accessible to a single particle. The canonical partition function is $Q(N,T,V)= \sum_{i} e^{-\beta H_{i}(N)}$ where $H_{i}(N)$ denotes the Hamiltonian of the $i^{th}$ state and $\beta=1/k_{b}T$ ($k_b$ is the Boltzmann constant). For the sake of simplicity, we address a homogeneous system of $N$ non-interacting identical particles in the volume $V$ (other than the fact that the states they can occupy are not independent one of each other). By defining $d_{N}$ as the number of states in $V$  accessible to the $N^{th}$ particle after $(N-1)$ have been added to $V$, then $Q(N,T,V)= W(N) e^{-\beta N U_{o}} q_{i}^{N}$ with \cite{haldane1991fractional}
	\begin{equation}
	W(N)= \frac{(d_{N}+N-1)!}{N! \ (d_{N}-1)!}
	\end{equation}
	where $U_{o}$ and $q_{i}$ are the energy per particle and the internal partition function, respectively. In the limit $n=\lim_{N,G \to \infty} N/G$, the thermodynamic functions are
	
	\begin{equation}{\label{eq.ftilde}}
	\begin{aligned}
	\beta \tilde{F}(n,T)&=\lim_{N,G \to \infty}\frac{F(N,T,V)}{G}=\lim_{N,G \to \infty}\frac{\ln Q(N,T,V)}{G}  \\
	&=\beta nU_{o}-[\tilde{d}(n)+n] \ln[\tilde{d}(n)+n] + \tilde{d}(n) \ln \tilde{d}(n)\\
	& \ \ + n \ln n
	\end{aligned}
	\end{equation}
	\begin{equation}{\label{eq.stilde}}
	\begin{aligned}
	\frac {\tilde{S}(n,T)}{k_{b}T}&=\lim_{N,G \to \infty}\frac{S(N,T,V)}{G}  \\
	&=[\tilde{d}(n)+n] \ln[\tilde{d}(n)+n] - \tilde{d}(n) \ln \tilde{d}(n) - n \ln n
    \end{aligned}
	\end{equation}
	and the chemical potential, $\mu=\left(\frac{\partial \tilde{F}}{\partial n}\right)_{T,V}$, satisfies
	\begin{equation}{\label{eqmu}}
	K(T) \ e^{\beta \mu}= \frac{n \ \left[ \tilde{d}(n) \right]^{\tilde{d}'(n)}}{\left[ \tilde{d}(n)+n\right] ^{\tilde{d}'(n)+1 }},
	\end{equation}
	where $\tilde{d}(n)=\lim_{N,G \to \infty} d_{N}/G$, $\tilde{d}'(n)= d[\tilde{d}(n) ]/dn$ and $K(T)=e^{-\beta U_{o}} \ q_{i}$.
	
	From Eq. \eqref{eqmu}, two related quantities are defined which will be later useful to fully interpret the state exclusion under spatial correlations. If the system of particles in $V$ is now assumed to exchange particles with a bath at chemical potential $\mu$ and temperature $T$, the time evolution of the state occupation $n$ is given by
	\begin{equation}{\label{eq.kinetic}}
	\frac{dn}{dt}= P_{o} \ W_{o \to \bullet}- P_{\bullet} \ W_{\bullet \to o},
	\end{equation}
	where $P_{o}(P_{\bullet})$ is the average fraction of empty (occupied) states in $V$ and $W_{o \to \bullet}(W_{\bullet \to o})$ the transition rate for an empty(occupied) state to get occupied (empty). In equilibrium, $dn/dt=0$, $W_{o \to \bullet}/W_{\bullet \to o}=P_{\bullet}/P_{o}=e^{\beta(\mu-U_{o})}$, $P_{\bullet}=n$. From Eq.\eqref{eqmu} and \eqref{eq.kinetic}
	\begin{equation}{\label{eq.Po}}
	P_{o}(n)=P_{\bullet}(n) \ e^{-\beta (\mu-U_{o})}= \frac{\left[ \tilde{d}(n)+n\right] ^{\tilde{d}'(n)+1 }}{\left[ \tilde{d}(n) \right]^{\tilde{d}'(n)} }.
	\end{equation}
	
	In addition, we introduce a new useful quantity, namely the exclusion spectrum function $\mathcal{G}(n)$, being the average number of excluded states per particle at occupation $n$ \cite{jjriccardo2018tesislic}. Thus, $\mathcal{G}(n)=\left\langle \frac{1}{N} \sum_{iº=1}^{G} e_{i} \right\rangle$
	\begin{equation}{\label{eq.Gn}}
	\begin{aligned}
	\mathcal{G}(n)&=%\left\langle \frac{1}{N} \sum_{iº=1}^{G} e_{i} \right\rangle
	\left\langle \frac{G}{N}\frac{1}{G} \sum_{i=1}^{G} e_{i} \right\rangle=\frac{1}{n}\left[ 1-P_{o}(n)\right]
	=\frac{1}{n}-\frac{1}{e^{\beta(\mu-U_{o})}}
	\end{aligned}
	\end{equation}
	where $e_{i}=1$ if the state $i$ out of $G$ is either occupied or excluded by any of the $N$ particles, or $e_{i}=0$ otherwise,  and the average is assumed to be taken over the canonical ensemble. The identity   $\left\langle\frac{1}{G}\sum_{i=1}^{G} e_{i} +P_{o}\right\rangle=1$ follows from the definition of $P_{o}$. $\mathcal{G}(n)$ characterizes the density dependence of the state exclusion for a spatially correlated many-particle system from zero-density to saturation.
	
	It is worth noticing that the rightmost side of Eq. \eqref{eq.Gn} also provides an operational formula to infer the exclusion spectrum $\mathcal{G}(n)$ from experiments. For instance, for adsorbed species under equilibrium conditions ($\mu,T$), $n$ is related to the surface coverage (so called adsorption isotherm) and  $U_{o}$ is obtained from the low density regime of $n(\mu,T)$.
	
	Spatially correlated states leading to multiple exclusion can be visualized, for instance,  in the   classical system of linear particles occupying sites on a square lattice (Fig. 1). Given the set of states for a single particle containing all its possible configurations on the lattice, clearly an isolated dimer ($C_{1}$) occupies one state plus excluding six more states from being occupied by other particles. For a larger number of particles on the lattice there exist configurations in which some states are excluded simultaneously by neighboring particles ($C_{2}$, $C_{3}$ and $C_{4}$). This is called here ``multiple exclusion" arising from spatial correlation between states, and it has significant effects on the thermodynamics of the system.
	\begin{figure}[h]
		\centering
		\includegraphics[width=1.00\columnwidth]{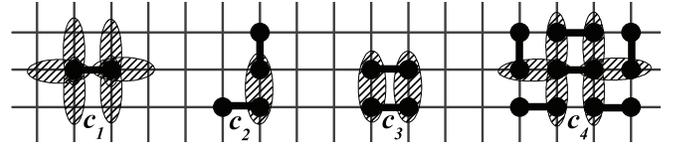}
		\caption{Local configurations of dimers on a square lattice. %showing multiple state exclusion (dashed).%
			$C_{1}$ shows the states (dashed) excluded by an isolated particle. $C_{2}$, $C_{3}$ and  $C_{4}$ depict states (dashed) multiply excluded by neighboring dimers, 1, 2 and 6 for $C_{3}$, $C_{2}$ and $C_{4}$, respectively.}
		\label{Fig.ejemplo.de.exclusion.k=2}
	\end{figure}

	It is known that the exact counting of configurations for an arbitrary number of particles on the lattice seems a hopeless task and it is still a relevant open problem in classical statistical mechanics. From here on, $d_{N}(\tilde{d}(n))$ is obtained through an approximation extending the Haldane-Wu's state counting procedure to a system of correlated states which determines the analytic multiple exclusion statistical distribution and the thermodynamics of the system. Given that the total number of states in $V$ is $G$, as we add particles from the $1^{st}$ to the $(N-1)^{th}$, the recursion relations can be written: $d_{1}=G$, $d_{2}=d_{1}-\mathcal{N}_{1},...,d_{N}=d_{N-1}-\mathcal{N}_{N-1}$, where $\mathcal{N}_{j}$ is the number of states occupied plus excluded only by the $j^{th}$ particle. Considering that a particle $j^{th}$ added to $V$ occupies one state and in addition it excludes a yet undetermined number of states out of $G$, we write the relation $\mathcal{N}_{j}=1+\mathcal{G}_{cj}$, where $\mathcal{G}_{cj}$ is the number of states excluded only by the $j^{th}$ particle [it does not account for the states excluded by $j$ which were already excluded by any of the particles $1,...,(j-1)$ because of the spatial correlations or so-called multiple state exclusion]. $\mathcal{G}_{cj}$ has to be rationalized as an average of over all the configurations of particles $1,....,j$ on the $G$ states. For $j \to N$ and $N,G \to \infty$ with $N/G=n$, it is straightforward that $\mathcal{G}_{cj}$ will converge to a value depending only on the ratio $N/G=n$ (as observed in simulation). Now we establish the following ansatz to determine $d_{N}$ \cite{jjriccardo2018tesislic}
	\begin{equation}{\label{eq.Nj}}
	\mathcal{N}_{j}=1+\mathcal{G}_{cj}=1+g_{c}\dfrac{d_{j}}{G},
	\end{equation}
	%where $g$ is the number of states excluded per particle assuming that each accessible state out of $G$ is independent of each other. As in Haldane's,  $g$ is the statistical exclusion parameter and it will given by the saturation $n_{m}=N_{m}/G=(G/g)/G=1/g$, being $n_{m}$ the maximum number of particles per state in $V$.%
	where $\mathcal{G}_{cj}=g_{c}\dfrac{d_{j}}{G}$, i.e, a system-dependent exclusion constant $g_{c}$ times the fraction $\dfrac{d_{j}}{G}$ of states that can be excluded  by particle $j$. It is worth mentioning that the second term in Eq. (\ref{eq.Nj}) resembles a sort of mean-field or effective-field approximation on the set of states which in the limit $N,G \to \infty$ will depend only on the mean occupation number $n=N/G$. Based on Eq. (\ref{eq.Nj}) we can rewrite the recursion relations as: $	d_{1}=G, d_{2}=d_{1}-\left[ 1+g_{c} \frac{d_{1}}{G} \right], d_{3}=d_{2}-\left[1+g_{c} \frac{d_{2}}{G} \right]=G\left[ 1-\frac{g_{c}}{G}\right]^{2}-\left[ 1-\frac{g_{c}}{G}\right]-1,...,d_{N}=d_{N-1}-\left[1+g_{c} \frac{d_{N-1}}{G} \right]=G \left[ 1-\frac{g_{c}}{G}\right]^{N-1}-\sum_{i=0}^{N-2} \left[ 1-\frac{g_{c}}{G}\right]^{i}$.

%	\begin{equation}{\label{eq.recurrence}}
%	\begin{aligned}
%	d_{1}&=G \\
%	d_{2}&=d_{1}-\left[ 1+g_{c} \frac{d_{1}}{G} \right]\\
%	d_{3}&=d_{2}-\left[1+g_{c} \frac{d_{2}}{G} \right]=G\left[ 1-\frac{g_{c}}{G}\right]^{2}-\left[ 1-\frac{g_{c}}{G}\right]-1\\
	%&.\\
	%&.\\
%	d_{N}&=d_{N-1}-\left[1+g_{c} \frac{d_{N-1}}{G} \right] \\
%	&=G \left[ 1-\frac{g_{c}}{G}\right]^{N-1}-\sum_{i=0}^{N-2} \left[ 1-\frac{g_{c}}{G}\right]^{i}\\
%	\end{aligned}
%	\end{equation}
	
	By taking the limit $\tilde{d}(n)=\lim_{N,G \to \infty}d_{N}/G$ it yields $\tilde{d}(n)=e^{-n g_{c}}-  n$. $\tilde{d}(n)$ is defined except for two constants, say $\tilde{d}(n)=C_{1} e^{-n g_{c}}-C_{2}  n$, provided that it must satisfy the boundary conditions $\tilde{d}(0)=1$ and $\tilde{d}(n_{m})=\tilde{d}(1/g)=0$, where the usual Haldane's exclusion constant $g$ is used here to denote the number of states excluded per particle at maximum occupation, $n_{m}=N_{m}/G=(G/g)/G=1/g$. Thus, $C_{1}=1$ and $C_{2}=g e^{-\frac{g_{c}}{g}}$ and finally
	\begin{equation}{\label{eq.dn}}
	\tilde{d}(n)=e^{-ng_{c}}-ge^{-\frac{g_{c}}{g}}n.
	\end{equation}
	
	We may even think of $g_{c}$ in Eq. \eqref{eq.Nj} as depending on $j$, i.e., $g_{cj}$. The recursion relations will lead to $d_{N}=d_{N-1}\left[1-g_{c(N-1)}/G \right]-1=G \prod_{j=1}^{N-1}\left[1-g_{cj}/G \right] - \sum_{i=2}^{N-1}\prod_{j=i}^{N-1} \left[1-g_{cj}/G \right]-1$. If $g_{cj}=g_{cN}+\Delta_{j,N}$, where $\Delta_{j,N}$ is finite, then $d_{N}=G{\left[1-g_{cN}/G \right]^{N-1}-\sum_{j=0}^{N-1}\left[1-g_{cN}/G \right]^{j}+\mathcal{O}(1/G)}$. In the $\lim_{N,G \to \infty}d_{N}/G$ it yields $\tilde{d}(n)=e^{-ng_{c}(n)}-n$ where $g_{c}(n)=\lim_{N,G \to \infty}g_{cN}$. From this, the ansatz \eqref{eq.Nj} is the simplest assumption on $g_{c}(n)$, $g_{c}(n)=g_{c}=$ constant, through which state exclusion is introduced in the state counting in presence of spatial correlations. This results in a fairly accurate approximation, as shown by comparing predicted observables and simulations for linear particle lattice gases.
	
	The exclusion constant $g_{c}$ is fully determined by the zero density limit of the mean number of states excluded particle, $\mathcal{G}(n)$. Accordingly, from Eqs.\eqref{eq.Po},\eqref{eq.Gn} and \eqref{eq.dn}
	\begin{equation}{\label{eq.Go}}
	\begin{aligned}
	\mathcal{G}_{o}=\lim_{n\to 0}\mathcal{G}(n)=\lim_{n\to 0} \left[1-P_{o}(n)\right]/n=2g e^{-g_{c}/g}+2g_{c}-1
	\end{aligned}
	\end{equation}
	$\mathcal{G}_{o}$ being the state exclusion at zero density, i.e, number of states excluded by an isolated particle in the system. Moreover, $\lim_{n\to n_{m}}\mathcal{G}(n)=\lim_{n\to n_{m}} \left[1-P_{o}(n)\right]/n=g$. The two exclusion constants, $g_{c}$ and $g$  in Eq. \eqref{eq.dn}, come from the infinite dilution and saturation limits of $\mathcal{G}(n)$, respectively.
	
	%Identical results (\eqref{eq.dn}) would be obtained upon satisfying the boundary conditions, if alternatively the ansatz of  eq. \eqref{eq.Nj} would be writen as $\mathcal{N}_{j}=g+g_{c} \ P_{j}$ %valid for $0<g<1$%
	%which reeds as: the states excluded by adding the particle $j$ to V equals  $g$ (characterizing its statistical exclusion in a set of independent states), plus $g_{c} \ P_{j}$ arising from particleÂ´s spatial  state correlations.
	
	%In general for a function $\mathcal{Z}{j}(d_{j})$ accounting for the number of states excluded by the particle $j$ [other than the ansatz of Eq. \eqref{eq.Nj}], then $N_{j}=1+\mathcal{Z}_{j}(d_{j})$ for $j=1,...,N$, $d_{N}=G-\sum_{i=1}^{N-1} \mathcal{Z}_{i}(d_{i})+(N-1)$ and $\tilde{d}(n)=\tilde{l}(n)-n$ where $\tilde{l}(n)=\lim_{N,G \to \infty}[1-\frac{1}{G} \sum_{i=1}^{N-1} \mathcal{Z}_{i}(d_{i})]$. Again, $\tilde{d}(n)$ will be completely determined by fulfilling the boundary conditions $\tilde{d}(0)$ and $\tilde{d}(n_{m})$.
	
	 From here on, we analyze linear $k$-mers ideal lattice gases under the proposed framework. We mean by linear $k$-mers, linear rigid particles made of $k$ identical beads occupying $k$ consecutive sites (one bead per site) on a regular lattice. For instance, this is a simple model for small polyatomics/hydrocarbons adlayers. For $k$-mers on a one-dimensional (1D) lattice, $g=k$, $\mathcal{G}_{o}=2k-1=2g-1$, the solution of Eq. \eqref{eq.Go} is $g_{c}=0$ $\forall k(\forall g)$ and the case reduces to Haldane's FE and Wu's distribution with $g=k$ resulting in the exact density dependence of the chemical potential $\mu\equiv\mu(n)_{T,V}$ from  Eq. \eqref{eqmu} (already derived in \cite{riccardo2004fractional} for non-interacting $k$-mers in 1D). In a $k$-mer 1D lattice gas, each state of $N$ $k$-mers on a lattice with $M=G$ sites and $n=N/M$  can be mapped onto a one of $N$ monomers on a equivalent lattice with $M'= M-(k-1)N$ sites and $n'=N/M'=n/[1-(g-1)n]$. Thus, there is not effective spatial correlation between excluded states for $k$-mers in 1D. On the other hand, for $k$-mers on a square lattice of $M$ sites, $G=2M$, $n_{m}=N_{m}/G=(M/k)/2M=1/(2k)=1/g$, then $g=2k$ and $\mathcal{G}_{o}=k^{2}+2k-1=\frac{g^{2}}{4}+g-1$. The solution of Eq. \eqref{eq.Go} is $g_{c}=\frac{g^{2}}{8}+\frac{g}{2}+g \mathcal{L}(z)$ for $g \geq 4$, where $\mathcal{L}(z)$ is the positive solution of $z=\mathcal{W}(z) e^{\mathcal{W}(z)}$, $ \mathcal{W}(z)$ being the Lambert function, namely, the inverse of $f(x)=x e^{x}, \ x=\mathcal{W}(x e^{x})$. Accordingly, $g_{c}=0$ for $k=2(g=4)$, $g_{c}=4.807$ for $k=3(g=6)$,            $g_{c}=9.586$ for $k=4(g=8)$,$g_{c}=15.344$ for $k=5(g=10)$,  $g_{c}=22.096$ for $k=6(g=12)$,
	$g_{c}=29.838$ for $k=7(g=14)$,  $g_{c}=38.563$ for $k=8(g=16)$, $g_{c}=48.267$ for $k=9(g=18)$,  $g_{c}=58.950$ for $k=10(g=20)$. Furthermore, $\lim_{k\to \infty}{g}_{c}=\mathcal G_{o}/2$.

	From Eq. \eqref{eqmu}, the occupation number, $n$, in general satisfies the following relation, formally almost identical to the transcendental equation first derived by Wu \cite{wu1994statistical}
	\begin{equation}{\label{distribution}}
	\left[\tilde{d}(n)+n\right]^{\tilde{d'}+1} \left[\tilde{d}(n)\right]^{-\tilde{d'}}=n \ e^{\beta\left(U_{o}-\mu \right) }=n \ \xi,
	\end{equation}
	where $\xi=e^{\beta\left(U_{o}-\mu \right) }$. From the explicit form of $\tilde{d}(n)$ [Eq. \eqref{eq.dn}], the distribution function can be symbolically written as
	\begin{equation}{\label{ME_distribution}}
	n=\frac{e^{-g_{c} n}}{w(\xi) + g \ e^{-g_{c}/g}},
	\end{equation}
	similar to Wu's distribution where $n\equiv n(\xi)$ is the solution of the transcendental Eq. \eqref{distribution} and $w(\xi)=\tilde{d}(n)/n$. For particles with exclusion parameter $g$ on spatially non-correlated states, $g_{c}=0$, $\tilde{d}(n)=1-gn$ and the Haldane's FE statistics is recovered and Eq. \eqref{ME_distribution} reduces to the Wu's distribution \cite{wu1994statistical}. Furthermore, $\tilde{d'}(n)=-g$ for $g_{c}=0$, thus $W(n)=\xi-1$ for $g=0$ and $w(n)=\xi$ for $g=1$, resulting Eq. \eqref{ME_distribution} the BE and FD statistics, respectively. Given that $w(n)=\tilde{d}(n)/n\geq0$, from Eq. \eqref{ME_distribution} the occupation-number's range is $0\leq n\leq 1/g$. At temperature $T=0$ (absolute scale), the distribution takes the step-like form $n=1/g$ for $U_{o}<\mu$ and $n=0$ for $U_{o}>\mu$, as expected.

\begin{figure}[h]
		\centering
		\includegraphics[trim={2.1cm 1.8cm 3.2cm 2.15cm},clip,scale=0.375]
		{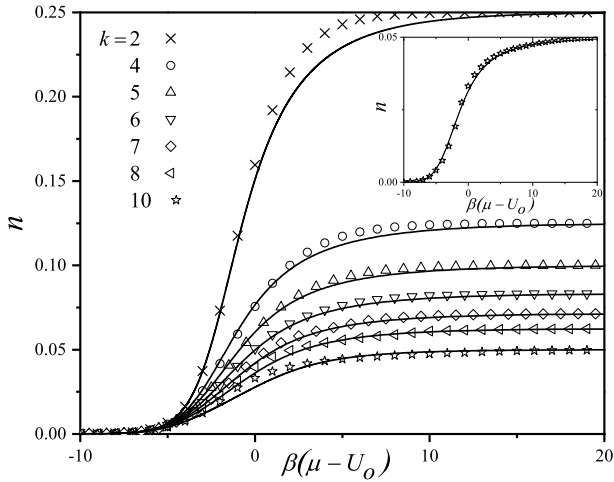}
		\caption{State occupation number $n$ versus $\beta(\mu-U_{o})$ for $k=2,4,5,6,7,8,10$ on a square lattice. Lines represent the analytical predictions from Eq. \eqref{eqmu}; symbols come from simulations. Inset shows the case $k=10$ for a smaller $g_{c}=39$ as to visualize the state exclusion effect of the nematic ordering.}
		\label{fig:nversusmu}
	\end{figure}

	Simulations of $k$-mers lattice gases were carried out in the Grand Canonical Ensemble through the efficient algorithm introduced by Kundu et al. \cite{kundu2013nematic,kunduprodeedings} to overcome the sampling slowdown at high density due to the jamming effects. The temperature, chemical potential $\beta \mu$ and system's size are held fixed and the number of particles on the lattice is allowed to fluctuate through non-local changes, i.e, insertion and deletion of $k$-mers at a time (in contrast to the standard Metropolis algorithm). Shortly, given a  configuration of $k$-mers on the lattice, one MCstep is fulfilled by removing all horizontal $k$-mers and keeping the vertical ones. The probabilities corresponding to horizontal segments of unoccupied sites are exactly calculated and stored for all the segment sizes. Then segments are occupied by $k$-mers with probabilities accordingly determined. An identical procedure is carried out in the vertical direction. A reproduction of these calculations is out of the scope of this work. The detailed discussion is found in the original work Refs.\cite{kundu2013nematic,kundu2014phase,kunduprodeedings}. The algorithm has proved to be ergodic, it satisfies the Detailed Balance Principle and equilibrium is reached after typically $10^{7}$ MC steps. $L \times L$ square lattices with periodic boundary conditions were used. The ratio $L/k$ was set to 120. With this value of $L/k$, we verified that finite size effects are negligible. The observables $\mathcal{G}(n)$ [Eq. \eqref{eq.Gn}] and $n=\left\langle N \right\rangle /G= \left\langle N \right\rangle/(2L^{2})$,  were calculated by averaging over $10^{7}$ configurations. The distribution function $n$ versus $\beta(\mu-U_{o})$ [Eq. \eqref{eqmu}]) is represented in Fig. \ref{fig:nversusmu} and compared with simulation for linear particles of size $k=2$ to $k=10$. %It was set $U_{o}=0$.
	
	The analytical predictions are accurate for all the particle sizes, being much better as $k$ increases up to $k=7$. The ansatz in Eq. \eqref{eq.Nj} does not account explicitly for system's dimensionality, shape or particles size and lattice structure, but all the state correlations are embedded in the exclusion constant $g_{c}$. For instance, the solid line in Fig. \ref{fig:nversusmu} for $k=2$ represents approximately the simulation results for dimers on the square lattice, $k=2 \ (\mathcal{G}_{o}=7,g=4)$, and it does exactly for tetramers on a 1D lattice, $k=4 \ (\mathcal{G}_{o}=7,g=4)$. For both cases the solution of Eq. \eqref{eq.Go} is $g_{c}=0$.
	
	 For $k\geq7$, it is known a nematic transition develops at intermediate lattice coverage with particles aligned along a lattice direction in compact clusters \cite{Ghosh}. Its effect is clearly seen in Fig.2 the case for $k=10$ at intermediate occupation where simulation and analytical function do not match. However, because the nematic ordering increases the number of multiply excluded states per particle, $n$ can be very accurately represented by the multiple exclusion statistics for a smaller value of the constant $g_{c}$ [according to the meaning of the corresponding term in Eq. \eqref{eq.Nj}] as shown in the inset of Fig. \ref{fig:nversusmu}.
	
\begin{figure}[h]
		\centering
		\includegraphics[trim={1.7cm 1.75cm 3.2cm 2.1cm},clip,scale=0.375]{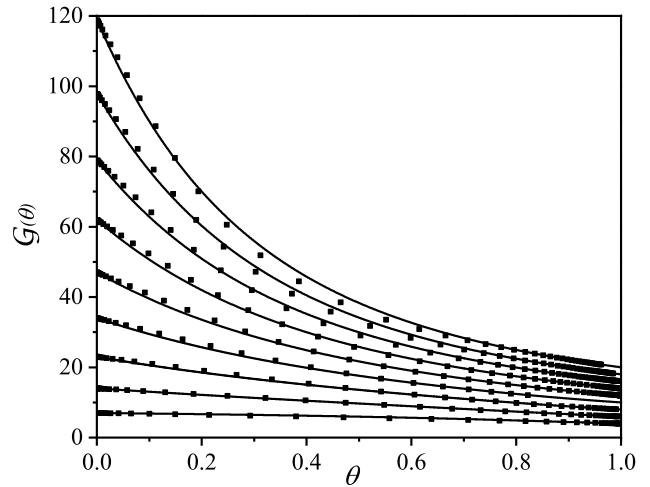}
		\caption{Exclusion spectrum $\mathcal{G}(\theta)$ for $k=2$ to $k=10$ (from bottom to top). Solid lines are analytical results from Eq. \eqref{eq.Gn} with $n=\theta/g=\theta/(2k)$. Symbols represent simulations.}
		\label{fig:gmedioversustitak2tok10}
	\end{figure}

	In addition, results for  the exclusion spectra $\mathcal{G}(n)$ from Eq. \eqref{eq.Gn}
	%and $P_{o}(n)$ from \eqref{eq.Po} and are shown in Figs.\ref{FigG} and \ref{FigPo},
	are shown in Fig. \ref{fig:gmedioversustitak2tok10} as a function of the lattice coverage $\theta=k<N>/M$, where $<N>$ and $M$ represent the average number of particles on the lattice and the number of lattice sites, respectively. Given that $\theta=k<N>/M=k<N>/(G/2)=2k<N>/G=g n$, all the quantities above can be expressed in the nomenclature of lattice coverage by the variable change $n=\theta/g$ with $0\leq\theta\leq 1$. The adsorption isotherm ($\mu$ vs $\theta$) follows straightforwardly from Eq. \eqref{eqmu} and \eqref{eq.dn}, $ \beta\mu=\ln[\frac{\theta}{g}]+[g_{c} e^{(-\theta g_{c}/g)}+ge^{(-g_{c}/g)}-1] \ln[e^{(-\theta g_{c}/g)}-e^{(-g_{c}/g)} \theta+\theta/g]- [g_{c} e^{(-\theta g_{c}/g)}+ge^{(-g_{c}/g)}] \ln[e^{(-\theta g_{c}/g)}-e^{(-g_{c}/g)} \theta]+\beta U_{o}$.
	%\begin{figure}[h]
	%	\centering
	%	\includegraphics[width=1\linewidth]{"../Julian Trabajo Final_TexStudio/Figuras Finales Julian/Po_versus_tita_k=2,3,5,7,10_PRL"}
	%	\caption{$P_{o}$ versus $\theta$. Solid lines from \eqref{eq.Po}, for $k=2,3,5,7$ and $10$ (from top to bottom). }
	%	\label{fig:poversustitak235710prl}
	%\end{figure}
	Concerning the new quantity we have introduced, $\mathcal{G}(\theta)$, the predictions from this work [Eq. \eqref{eq.Gn} along with  \eqref{eq.Po} and \eqref{eq.dn}] reproduce significantly well the exclusion per particle for all $k$ as density varies. This appears as a very useful function in the presence of correlations since can be obtained directly either from the distribution $n(\mu)$ or from experiments providing a relevant average measurement about the spatial configuration of particles in the system from thermodynamics. The limiting values being  $\mathcal{G}(0)=\mathcal{G}_{o}$ and $\mathcal{G}(1)=g$. Additionally, state exclusion can be observed through $\mathcal{G}(\theta)$ in the presence of particle interactions and order-disorder transitions, as it will be presented in future work.
	Finally, an approach to the equilibrium statistics of many-particle systems with exclusion having spatially correlated states for single-particles has been put forward, the statistical distribution has been obtained, a useful exclusion spectrum function has been defined and the results applied to 2D-lattices from small to large linear particles, resulting in a significant agreement for such a complex statistical systems. The formalism can be straightforwardly applied to other particles/lattice geometries and higher dimensions. In addition, the analysis could be extended to more complex off-lattice systems in the presence of mutual exclusion (such as hard disks and spheres in the continuum). This work is in progress.

This paper was supported in part by CONICET and Universidad Nacional de San Luis, Argentina.

\end{document}